# A puzzle of radio-quiet quasar PG 1407+265: are optical and X-ray emissions produced from relativistic jet?

Liang Chen[1,2,3] and Jinming Bai[1,2]

[1] *National Astronomical Observatories/Yunnan Observatory, Chinese Academy of Sciences, Kunming 650011, China;*
[2] *Key Laboratory for the Structure and Evolution of Celestial Objects, Chinese Academy of Sciences;*
[3] *Graduate School of Chinese Academy of Sciences;*

PG 1407+265 is a radio quiet quasar but has a relativistic jet. In this report, we show some peculiar properties of its optical and X-ray emissions, which indicate their possible non-thermal origins produced from jet. We use a simple synchrotron + synchrotron self Compton (SSC) model to fit the emissions with different ratios of energy densities between magnetic field and electrons ($\eta \equiv U_B / U_e$), which predict different γ-ray luminosity. The First LAT AGN Catalog (1LAC) did not include PG 1407+265, which indicates an upper limit of γ-ray luminosity. This upper limit constrains the ratio unreasonable large ($\eta \geq 10^{4-5}$). This inversely indicates that the optical and X-ray emissions may be not produced from the beaming jet. We discuss the physical implications of these results.

## 1 Introduction

Kellermann et al. [1] showed that in optically selected sample, quasars with similar optical properties can have very different radio properties. The quasars form a dichotomy in radio loudness (R) distribution (i.e., radio-loud, RL and radio-quiet, RQ). While the study on FIRST quasars showed that the radio-loudness distribution is not bimodal but continuous [2]. The debate on the validity of the bimodal or continuous distributions is ongoing [2-4]. It is believed that RL AGNs have relativistic jets, while RQ AGNs have no or very weak jets. VLBI observations reveal more and more RQ/radio-intermediate AGNs having jet structure [5-7] and some have knots being relativistic (e.g., PG 1407+265 [8] and III Zw 2 [9]).

The study on the AGN from RQ to RL can offer important information on our understanding on the growth of jet, and the interactions between jet and environment [10]. In the unified scheme of RL AGNs [11], blazars having relativistic jets point to earth. The SEDs show two bumps, peaking at infrared (IR) to X-ray bands and at γ-ray band, respectively [11]. The lower bump is thought to be the synchrotron emissions of high energy electrons. The high one may be the inverse Compton (IC) emissions by the same electron population ([12-15]; see also the hadronic model [16-19]).

PG 1407+265 may evolve to be an important laboratory to investigate the relationship between RQ and RL AGNs, because the quasar has both RL and RQ properties: a RQ AGN but launching a relativistic radio jet [20]. In this report, we study its optical and X-ray emissions through modeling the SED with synchrotron + synchrotron self Compton (SSC) model. Throughout the report, a cosmology

Corresponding authors (email: chenliangew@hotmail.com; baijinming@ynao.ac.cn )

with $H_0 = 70 \text{ km s}^{-1} \text{ Mpc}^{-1}$, $\Omega_m = 0.3$ and $\Omega_\Lambda = 0.7$ is adopted.

## 2  General properties of PG 1407+265

PG 1407+265 is a RQ AGN (R=3.48, $z$=0.94; [1,21]) with unusual properties. From radio morphology and variations, Blundell et al. [8] concluded that the quasar holds a relativistic jet pointing to earth with beaming factor $\delta \geqslant 10$. The optical and X-ray emissions of PG 1407+265 show some peculiar properties. These apparent unusual properties can be understood if we assume that they are non-thermal emissions from the beaming jet. McDowell et al. [21] showed that the lines do not catch up the UV continuum variability, which indicates that the continuum emission may not come from the central accretion disk, thus it maybe from the beaming jet. In BL Lac objects, the emission lines are shined off by the beaming continuum emissions, while we can look directly into the inner part of the broad lime region (BLR), if they have. The equivalent widths (EW) of emission lines of PG 1407+265 are lower 3-10 times than typical values of quasars, while the velocity dispersion of $\upsilon \approx 10000 \text{ km s}^{-1}$ is relatively larger [21]. This can be explained if the continuum emission is the jet beaming emission as that in BL Lacs. The optical-to-X-ray spectrum index ($\alpha_{oX}$) of typical RQ AGNs ranges from 1.2 to 1.8, and the slope steepens as increasing UV luminosity [22]. During the high state of PG 1407+265, the optical-to-X-ray spectrum is flat ($\alpha_{oX}$ =1.09) and there are higher fluctuations in X-ray than UV bands (i.e., the slope flatting with increasing luminosity; [20]), which implies different radiation mechanisms in PG 1407+265 from typical RQ AGNs. Gallo [20] suggested that the soft X-ray may be the jet non-thermal emission during the high state. During this high state, UV variability correlates with that of X-ray [20]. This also indicates that the emissions from optical to X-ray bands may come from the jet. Another indirect evidence is that the radiation efficiency derived from X-ray variability ($\eta_{eff} \geqslant 0.29$, isotropy assumed) is comparable to that in maximally rotating Kerr black hole ($\eta_{eff} \approx 0.3$; [23]). This extreme demand can be relaxed if adopted relativistic beaming emission [20].

These properties indicate that the optical and X-ray emissions may come from the jet. We collect the broad bands SED of PG 1407+265 from *NED* (blue dots in Figure 1). The red triangle (with the bow-tie) is for the low state X-ray observation from [24]. In next section, we model its optical and X-ray emissions by using a simple synchrotron + SSC model.

## 3  SED modeling

We assume that the emission region is a homogeneous sphere with radius $R$ embedded in magnetic field $B$ [25,26]. The electron energy distribution is assumed to be a break power law,

$$N(\gamma) = \begin{cases} N_0 \gamma^{-p_1} & \gamma \leqslant \gamma_p \\ N_0 \gamma_p^{p_2-p_1} \gamma^{-p_2} & \gamma > \gamma_p \end{cases}, \quad (1)$$

where $\gamma_p$ is the peak electron energy, $N_0$ is the normalized density, and the indexes $p_{1,2} = 2\alpha_{1,2} + 1$, $\alpha_{1,2}$ are the spectral indexes below and above the peak frequency.

From the Figure 1, it can be seen that the SED of PG 1407+265 has a bump at IR band. This may be the emissions from the torus, which beyond the scope of this report. Radio emissions come from more extended regions relative the higher frequency emissions, which cannot be modeled in our model. It also can be seen that the SED may have a peak between optical and X-ray band, and this component may be the synchrotron emissions. Blundell et al. [8] derived the lower limit of the beaming factor $\delta \geqslant 10$ for PG 1407+265. In our SED modeling, we take this lower limit $\delta$=10. The size of the emission is constrained by the minimal variability timescale $R \sim t_{var} c\delta/(1+z) = 1.5 \times 10^{16}$ cm [20]. Very hard spectrum (inverse spectrum) at optical band constrains $p_1 = 2\alpha_1 + 1 = 0.6$. This so hard index makes that electrons at peak energy contribute the main electron energy density ($U_e \approx N_0 mc^2 \gamma_p^{2-p_1}/(2-p_1)$, see Equation 1). $p_2 = 4.2$ is taken to describe the high energy branch of the electron energy distribution and the reasonable alternatives to this values do not affect our conclusions.

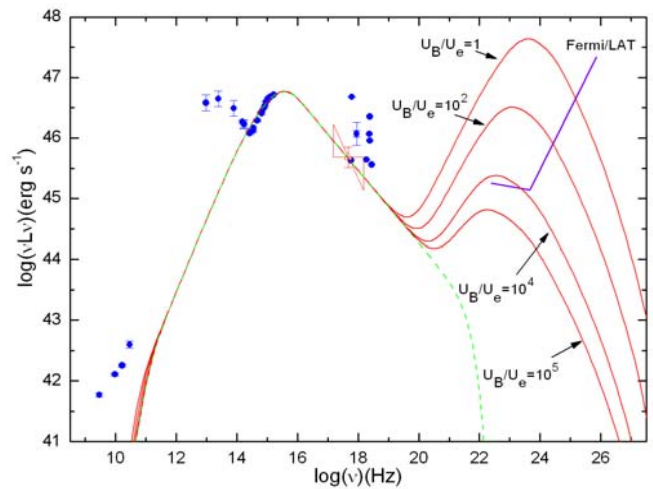

**Figure 1.** The calculated SEDs with different ME ratios (red solid lines, parameters are listed in Table 1). The synchrotron emission is presented as the green dashed line. The violet line is the sensitivity of detector *Fermi*/LAT [28]. The blue dots are the observational data from *NED*. The red triangle (with the bow-tie) is for the low state X-ray observation from [24].

For the reason of lacking the IC component, we need an extra constraint in addition to the SED to determine other jet parameters. Here we introduce a parameter: the ratio of energy densities between magnetic field and electrons in the jet frame ($\eta \equiv U_B/U_e$, hereafter ME ratio). The ME ratio has typical values around 10 [27,28]. Modeling with different ME ratios produce the similar synchrotron emissions, but would predict different IC emissions. With the ME ratio, all jet parameters can be determined [25]. In the IC calculation, we only consider SSC. The external Compton (EC) emissions will certainly increase the γ-ray emissions, and this does not affect our conclusions.

**Table 1.** Jet main parameters with different ME ratios

| $U_B/U_e$ | $\delta$ | $R$ (cm) | $B$ (Gs) | $N_0$ | $\gamma_p$ | $p_1$ | $p_2$ |
|---|---|---|---|---|---|---|---|
| 1 | 10 | $1.5\times10^{16}$ | 1.29 | 0.45 | 7215 | 0.6 | 4.2 |
| $10^2$ | 10 | $1.5\times10^{16}$ | 4.82 | 0.16 | 3737 | 0.6 | 4.2 |
| $10^4$ | 10 | $1.5\times10^{16}$ | 18.0 | 0.055 | 1935 | 0.6 | 4.2 |
| $10^5$ | 10 | $1.5\times10^{16}$ | 34.7 | 0.032 | 1393 | 0.6 | 4.2 |

## 4  Results and discussion

The calculated SEDs are showed in Figure 1. In which we model the X-ray emissions at low state [24]. The modeling of the high state will predict higher γ-ray emission, and this does not affect our conclusions. We can see that the γ-ray luminosity decrease with increasing ME ratios. We also present the sensitivity of detector *Fermi*/LAT [29]. It can be seen that only the extremely large ME ratio can result in the γ-ray flux lower than the sensitivity of *Fermi*/LAT. *Fermi*'s main observing mode is the sky-survey mode. In this mode, the LAT observes the entire sky every 3 hr [30]. Abdo et al. [30] presented the first LAT AGN Catalog (1LAC) in which the data were collected from Aug. 4 2008 to Jul. 4 2009, primarily with sky-survey observations. We note that PG 1407+265 is not included in 1LAC. This indicates that the ME ratio must be unreasonably large if the optical and X-ray emissions come from the beaming jet ($\eta \geq 10^{4-5}$, [27,28]).

Gallo [20] suggested that the X-ray emission is combination of two components: a hard one from the accretion disk and a steep one from the beaming jet. Optical data lacks the polarization and do not show violent variability [21,31]. These indicate that the optical emissions may come from the accretion disk. In this case, the electron distribution would not be constrained. Hence, the constraint on the ME ratio should be relaxed. Some studies indicated that the optical and X-ray emissions should not be ignored in some radio-loud AGNs [32-34]. We take the optical luminosity as bolometric one $10^{46-47}$ erg s$^{-1}$ and the full width at half-maximum (FWHM) of emission line Mg II 7000 km s$^{-1}$ [21] to estimate the black hole mass and the Eddington ratio ($\eta$) of PG 1407+265 (e.g., [35] and references therein). The estimations give $M_\bullet \approx (2.5-8)\times10^9 M_\odot$ and $\eta \approx 0.03-0.1$. This is consistent with the standard disk model [36], which indicates that the optical emissions may be thermal emissions from accretion disk.

On the other hand, there are also some properties can not be understood well. McDowell et al. [21] showed that when the UV continuum brightens, the emission lines had not yet caught up. This may be explained if an effective radius of BLR is large compared with 10 lt-yr. But deduced from the relation between the luminosity and the radius, the radius of BLR is only ~1 lt-yr [21]. The nature of the weak lines emission is also unclear if the optical and UV emissions come from accretion disk [20]. From Fig. 6 in Gallo [20], it can be seen that the variability of optics correlates with that of X-ray and possibly leads the X-ray with about several kilo seconds during the high state. During this high state, the soft X-ray dominates the 0.25-10 keV emission, and the soft X-ray emission leads the hard one about 3000 s. If optics and hard X-ray come from the accretion disk and the soft X-ray comes from the jet, this needs fine geometry of the emission region to explain the above phenomena: optics leading soft X-ray which leads hard X-ray.

As a summary, some properties of PG 1407+265 show that the optical and X-ray emissions may come from the beaming jet. The fact that its γ-ray emission has not been detected by *Fermi*/LAT requires an unreasonable large ME ratio. This is inversely to suggest that the optical and X-ray emissions may not all produced from the jet. The nature of PG 1407+265 remains puzzling. The detailed simultaneous SED from radio to X-ray bands and the VLBI observation with higher resolution are needed to unveil the nature of PG 1407+265.

*We thank the anonymous referees for helpful comments and suggestions. L.C. thanks the West PhD project of the Training Programme for the Talents of West Light Foundation of the CAS, and National Natural Science Foundation of China (NSFC; Grant 10903025 and 10778702) for financial support. J.M.B. thanks support of NSFC (Grant 10973034) and the 973 Program (Grant 2009CB824800).*